\documentclass[twocolumn,showpacs,amsmath,amssymb]{revtex4}
\usepackage{graphicx} 
\usepackage{natbib} \usepackage{fancyhdr} \usepackage{color}
\usepackage{amsmath} \usepackage{epsfig} \usepackage{multirow}
\usepackage{amssymb} \usepackage{algorithm,algorithmic}

\begin{document}

\title{Generalized Master Equations for Non-Poisson Dynamics on Networks}
\author{Till Hoffmann} \affiliation{Department of Physics, University of Oxford, Oxford, United Kingdom} 
\author{Mason A. Porter} \affiliation{Oxford Centre for Industrial and Applied Mathematics, Mathematical Institute, University of Oxford, Oxford, OX1 3LB, United Kingdom} \affiliation{CABDyN Complexity Centre, University of Oxford, Oxford OX1 1HP, United Kingdom} 
\author{Renaud Lambiotte}\email[Email: ]{renaud.lambiotte@fundp.ac.be} \affiliation{Naxys, University of Namur, B-5000 Namur, Belgium} \affiliation{Department of Mathematics, University of Namur, B-5000 Namur, Belgium}

\begin{abstract}
The traditional way of studying temporal networks is to aggregate the dynamics of the edges to create a static weighted network. This implicitly assumes that the edges are governed by Poisson processes, which is not typically the case in empirical temporal networks.  Consequently, we examine the effects of non-Poisson inter-event statistics on the dynamics of edges, and we apply the concept of a generalized master equation to the study of continuous-time random walks on networks. We show that the equation reduces to the standard rate equations when the underlying process is Poisson and that the stationary solution is determined by an effective transition matrix whose leading eigenvector is easy to calculate. We discuss the implications of our work for dynamical processes on temporal networks and for the construction of network diagnostics that take into account their nontrivial stochastic nature.
\end{abstract}


\date{\today}

\maketitle

\section{Introduction}

Over the past two decades, myriad papers have illustrated that understanding of complex systems composed of large numbers of interacting entities can be improved considerably by adopting a network-science perspective \cite{review}. In particular, the influence of network architecture on dynamical processes such as random walks, biological and social epidemics, and opinion formation is now well-appreciated (and somewhat better understood than it used to be) \cite{durrett2010}. The effects of network structure on dynamics on networks have been investigated using a variety of techniques, including mean-field theories and pair approximations, spectral methods, and numerical simulations \cite{boccaletti}.

In the study of dynamics on networks, there has been intense focus on network structure---i.e., the arrangement of edges and their associated weights---but the effects of the temporal patterns of edges remains very poorly understood \cite{reviewSaram}. In a realistic setting, many networks are not static, as both edges and nodes can appear and disappear, and this has a strong effect on spreading processes \cite{importanceofdynamics0,importanceofdynamics,Moro,tang2010,rocha,karsai,takaguchi,maxi}. Most studies of networks tend to overlook such temporal patterns by assuming a time-independent set of nodes and assigning a single scalar to represent a weighted edge $A_{ij}$ between each pair of nodes $i$ and $j$. This scalar is supposed to represent some sort of aggregate importance of the connection between $i$ and $j$, and it is often understood as a rate in the case of continuous-time processes \cite{vazquez2}.  From such a Poisson perspective, the probability that an edge appears between two nodes $i$ and $j$ in a time interval of length $dt$ is given by $dt/\langle \tau_{ij} \rangle$, where $\langle \tau_{ij} \rangle \equiv 1/A_{ij}$ (where $\tau_{ij} = +\infty$ if $i$ and $j$ are not connected by an edge) is the mean inter-event time and the time $\tau$ between two consecutive events obeys an exponential distribution with mean $\langle \tau \rangle$.

This Poisson assumption facilitates theoretical analysis, provides intuitive results, and is reasonably accurate for a variety of physical and biological systems in which the rate at which events take place does not depend on their histories. However, numerous socio-economic systems exhibit non-Poisson temporal statistics due to their non-stationarity or their non-Markovian nature. Observations made in financial markets \cite{sabatelli2002waiting}, email \cite{burstystreams,Eckmann,bara,malmgren}, letter \cite{oliveira} or online \cite{burstystreams2,inform} communication networks, and face-to-face contacts \cite{isella}, have shown that the time intervals between isolated actions performed by an individual or between isolated interactions for a pair of individuals deviates significantly from a Poisson process. This behavior has important implications, e.g. for the spreading of epidemics \cite{caley2007influenza}.

Two perspectives have been used to attempt to go beyond the unrealistic Poisson assumption. First, one can perform simulations on temporal graphs for which a time series of the presence versus absence of edges is deduced directly from empirical observations \cite{rocha,karsai}. However, such a computational approach has a significant drawback: It relies entirely on simulations and/or algorithms to produce, e.g., time-randomized versions of the original data, and it is thus unable to improve underlying theoretical understanding or build predictive models of the problem. Second, one can study the problem in a more abstract manner by developing spreading models that incorporate realistic temporal statistics. Such models can then be studied either mathematically or using numerical simulations \cite{takaguchi,vazquez2}. In this second approach, the underlying network is seen as a fluctuating entity. One typically assumes that it is driven by a stationary stochastic process, which allows exhaustive computational experimentation through the analysis of ensembles of realizations and by tuning structural and temporal parameters. This modeling perspective also lends itself to mathematical analysis \cite{new1,new2}, though most research on it thus far has been computational in nature.

In this paper, we take the second approach and argue for the development of a mathematical framework to explore the effect of non-Poisson inter-event statistics on dynamics. To do so, we apply the concept of a generalized master equation \cite{montroll1965}, which is traditionally defined on networks with regular structures (i.e., on regular lattices), to the study of continuous-time random walks on networks. Generalized master equations lie at the heart of the theory for anomalous diffusion and have applications ranging from ecology \cite{ap2} to transport in materials \cite{ap1}. Our choice regarding what dynamics to consider is motivated by the importance of random walks as a way to understand how network structure affects dynamics and to uncover prominent structural features from networks.

The rest of this paper is organized as follows. We first introduce the process and derive its generalized master equation in both ordinary space and Laplace space. After checking that the equation reduces to standard rate equations when the underlying process is Poisson, we focus on its stationary solution and show that it is determined by an effective transition matrix whose leading eigenvector can be calculated rapidly even in very large networks. Finally, we discuss the implications of our work in terms of dynamical processes on time-dependent networks and for the construction of network diagnostics that take into account their nontrivial stochastic nature.


\section{Generalized Master Equation}

In this section, we derive a generalized master equation for continuous-time random walks with arbitrary waiting-time statistics.


\subsection{Beyond Static Weights on Edges}

\begin{figure} 
\begin{centering}
\includegraphics[width=0.4\textwidth]{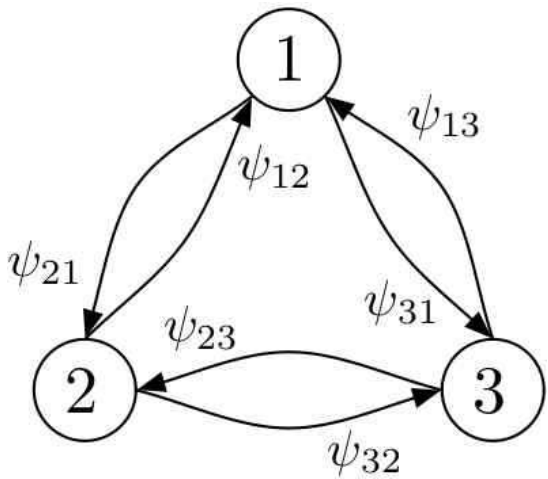} \par
\end{centering}
\caption{\label{fig:graph}A directed graph with $N=3$ nodes and no self-loops. The waiting-time distribution $\psi_{ij}\left(t\right)$ characterizes the appearances of an edge from $j$ to $i$.} 
\end{figure}

Consider the $N$-node directed graph $\mathcal{G}$ in Fig.~\ref{fig:graph}. To avoid unnecessary complications, we assume for this discussion that $\mathcal{G}$ is strongly connected (which means that one can reach any node starting from any other node).  A network is typically represented using an adjacency matrix ${A}$ whose elements $A_{ij}$ indicate the connection strength between each pair of nodes $i$ and $j$. For an unweighted network, $A_{ij}$ can take either the value 1 (if there is an edge) or the value 0 (if there is not one). For a weighted network,  $A_{ij}$ can take an arbitrary value (which is also non-negative in most studies), and a value of $0$ again indicates the absence of an edge. 

Whether one is considering a weighted or an unweighted network, one typically aggregates networks evolving over a certain time interval and thus ignores the temporal pattern of edges, which can be of critical importance in empirical situations \cite{reviewSaram}. We propose instead to assign to each edge an inter-event time distribution that determines when a edge is accessible for transport. The dynamics of a network are thus characterized by an $N\times N$ matrix $\psi\left(t\right)$ of smooth, piecewise continuous waiting-time distributions (WTDs) $\psi_{ij}\left(t\right)$ that determine the appearances of an edge emanating from node $j$ and arriving at node $i$. We further assume that the edges remain present for infinitesimally small times, which implies that the network is empty (and thus that the entries of the adjacency matrix are $0$), except at random instantaneous times determined by $\psi\left(t\right)$, when a single edge is present. From now on, we use the following terminology: $\mathcal{G}$ is called the underlying graph---i.e., this graph determines which edges are allowed and which are not; and $\psi_{ij}\left(t\right)$ determines a dynamical graph, in which edges appear randomly according to the assigned waiting times. We use this random process to model the transitions of a random walker moving on the graph. By construction, a walker located at a node $j$ remains on it until an edge leaving $j$ toward some node $i$ appears. When such an event occurs, the walker jumps to $i$ and waits until an edge leaving $i$ appears.

There are several ways that one can set up the WTDs $\psi_{ij}\left(t\right)$. We consider the case in which a WTD corresponds to the probability for an edge to occur between time $t$ and $t+dt$ after the random walker arrives on node $j$ in the previous jump. In other words, all edges leaving $j$ have their clocks reinitialized when a walker arrives on it \footnote{Another possibility is for the appearances of edges around a node according to independent processes.}.  It follows from the definition that
\[ 
	\int_{0}^{\infty} \psi_{ij}\left(t\right)=1\,. 
\] 
The probability that an edge appears between $j$ and $i$ before time $t$ is 
\[
	\int_{0}^{t}\psi_{ij}\left(t'\right)dt'\,, 
\] 
and the probability that it does not appear before time $t$ is therefore
\[
	\chi_{ij}\left(t\right)\equiv1-\int_{0}^{t}\psi_{ij}\left(t'\right)dt'\,. 
\] 
If a transition from $j$ to $i$ is not allowed, then the corresponding element $\psi_{ij}(t)$ is equal to $0$ for all times. Additionally, we assume for simplicity that the underlying network $\mathcal{G}$ has no self-loops.

It is important to distinguish between the WTD $\psi_{ij}\left(t\right)$ of the process that might lead to a step and the probability distribution $T_{ij}\left(t\right)$ for actually making a step to $i$. This distinction is necessary because all of the processes on a node are assumed to be independent of one another, but the probability to make a step depends on all of the processes. As an illustration, consider a walker on a node $j$ with only one outgoing edge to $i$. The probability distribution function (PDF) to make a step to $i$ a time $t$ after having arrived on $j$ is then 
\[
	T_{ij}\left(t\right)=\psi_{ij}\left(t\right)\,. 
\] 
However, if there exists another edge leaving $j$ (e.g., an edge to node $k$), then the PDF to make a transition to $i$ is modified, as a step to $i$ only occurs if the edge to $i$ appears before the one going to $k$. In this situation, 
\[
	T_{ij}\left(t\right)=\psi_{ij}\left(t\right)\chi_{kj}\left(t\right)\,. 
\] 

In general, the PDF to make a step from $j$ to $i$ accounting for all other processes on $j$ is
\begin{align} 
	T_{ij}\left(t\right) & =\psi_{ij}\left(t\right)\times\prod_{k\neq i}\chi_{kj}\left(t\right) \nonumber \\ 
	& =\psi_{ij}\left(t\right)\times\prod_{k\neq i}\left(1-\int_{0}^{t}\psi_{kj}\left(t'\right)dt'\right)\,.
	\label{eq:T_ij(t)} 
\end{align} 
Equation (\ref{eq:T_ij(t)}) emphasizes the importance of the temporal ordering of the edges in the diffusive process. It can also be written as 
\begin{equation}
	T_{ij}\left(t\right)=-\frac{d\chi_{ij}\left(t\right)}{dt}\times\prod_{k\neq i}\chi_{kj}\left(t\right)\,,
	\label{eq:T_ij(t)-differential-form} 
\end{equation}
as $\chi'_{ij}(t)=-\psi_{ij}\left(t\right)$. An extreme scenario occurs when $\psi_{ij}=\delta(t-t_{ij})$ and $\psi_{kj}=\delta(t-t_{kj})$ for $t_{ij}>t_{kj}$.  In this case, $T_{ij}(t)=0$, so a walker located on $j$ never takes the edge to $i$ even if this edge appears very frequently as a function of time.


\subsection{Generalized Montroll-Weiss Equation}

We now focus on the trajectories of a random walker exploring a temporal, stochastic network. We closely follow the standard derivation of the Montroll-Weiss (MW) equation \cite{montroll1965} and generalize it to an arbitrary $N$-node network of transitions. The probability $n_{i}\left(t\right)$ to find a walker on node $i$ at time $t$ is the integral over all probabilities $q_{i}\left(t'\right)$ of having arrived on $i$ at time $t'<t$, weighted by the probability $\phi_{i}\left(t-t'\right)$ of not having left the node since then:
\begin{equation}
	n_{i}\left(t\right)=\int_{0}^{t}\phi_{i}\left(t-t'\right)q_{i}\left(t'\right)dt'\,.
	\label{eq:n_i(t)} 
\end{equation} 
We take the Laplace transform $\hat{n}_{i}\left(s\right)\equiv\mathcal{L}\left\{ n_{i}\left(t\right)\right\}=\int_{0}^{\infty}n_{i}\left(t\right)e^{-st}dt$ to exploit the fact that the convolution reduces to a product in Laplace space \cite{dyke1999introduction}:
\begin{equation}
	\hat{n}_{i}\left(s\right)=\hat{\phi}_{i}\left(s\right)\hat{q}_{i}\left(s\right)\,.
	\label{eq:n_i(s)}
\end{equation}

We obtain the quantity $\hat{\phi}_{i}\left(s\right)$ from the PDF as follows. Making a step from node $i$ to any other node gives 
\begin{equation}
	T_{i}\left(t\right)=\sum_{j=1}^{N}T_{ji}\left(t\right)\,. 
	\label{eq:T_i(t)}
\end{equation} 
(Note that $T_{ii} = 0$ because there are no self-loops.) The PDF to remain immobile on $i$ for a time $t$ is thus 
\begin{align} 
	\phi_{i} & =1-\int_{0}^{t}T_{i}\left(t'\right)dt'\,. 
	\label{phiphi}
\end{align} 
Taking the Laplace transform of equation (\ref{phiphi}) then yields 
\begin{equation}
	\hat{\phi}_{i}\left(s\right)=\frac{1}{s}\left(1-\hat{T}_{i}\left(s\right)\right)\,.
	\label{eq:phi_i(s)} 
\end{equation}

The quantity $\hat{q}_{i}\left(s\right)$ is the Laplace transform of the PDF $q_{i}\left(t\right)$ to arrive on node $i$ exactly at time $t$. One calculates it by accounting for all $k$-step processes that can lead to such an event \cite{scher1973}: 
\[
	q_{i}\left(t\right)\equiv\sum_{k=0}^{\infty}q_{i}^{\left(k\right)}\left(t\right)\,,
\] 
where $q_{i}^{\left(k\right)}\left(t\right)$ represents the probability to arrive on node $i$ at time $t$ in exactly $k$ steps. Note that the PDF at $i$ is related to that at $j$ by the recursion relation 
\begin{equation} 
	q_{i}^{\left(k+1\right)}\left(t\right)
	=\int_{0}^{t}d\tau\sum_{j}T_{ij}\left(t-\tau\right)q_{j}^{\left(k\right)}\left(\tau\right)\,.
	\label{steps}
\end{equation} 
In other words, the probability to arrive on node $i$ in $k+1$ steps is the probability to arrive on any other node $j$ in $k$ steps weighted by the probability of making a step $j\rightarrow i$ at the required time. 

Upon taking a Laplace transform, equation (\ref{steps}) becomes 
\[
	\hat{q}_{i}^{\left(k+1\right)}\left(s\right)=\sum_{j}\hat{T}_{ij}\left(s\right)\hat{q}_{j}^{\left(k\right)}\left(s\right)\,.
\] 
Summing over all $k$ and adding $\hat{q}_{i}^{\left(0\right)}\left(s\right)$ yields 
\begin{align*}
	\hat{q}_{i}^{\left(0\right)}\left(s\right)+\sum_{k=0}^{\infty}\hat{q}_{i}^{\left(k+1\right)}\left(s\right)
	&=\sum_{j}\hat{T}_{ij}\left(s\right)\sum_{k=0}^{\infty}\hat{q}_{j}^{\left(k\right)}\left(s\right)+\hat{q}_{i}^{\left(0\right)}\left(s\right)\,,
\end{align*} 
which can also be written in terms of matrices and vectors: 
\begin{align*}
	\hat{{q}}\left(s\right) 
	&=\hat{T}\left(s\right)\hat{{q}}\left(s\right)+\hat{{q}}^{\left(0\right)}\left(s\right)\,.
\end{align*} 
Noting that ${q}^{\left(0\right)}\left(t\right)={n}\left(0\right)\delta\left(t\right)$, we see that the last term is simply ${n}\left(0\right)$, which leads to the following solution in Laplace space: 
\begin{equation}
	\hat{{q}}\left(s\right)=\left(I-\hat{T}\left(s\right)\right)^{-1}{n}\left(0\right)\,.
	\label{eq:q(s)} 
\end{equation}

We insert the expression (\ref{eq:phi_i(s)}) for $\hat{\phi}_{i}\left(s\right)$ and the expression (\ref{eq:q(s)}) for $\hat{q}_{i}\left(s\right)$ into the equation for the walker density (\ref{eq:n_i(s)}) to obtain a generalization of the MW equation \cite{montroll1965} that applies to arbitrary network structures: 
\begin{align} 
	\hat{n}_{i}\left(s\right) 
	&=\frac{1}{s}\left(1-\hat{T}_{i}\left(s\right)\right)\sum_{k}\left(I-\hat{T}\left(s\right)\right)_{ik}^{-1}n_{k}\left(0\right)\nonumber \\ 
	&=\sum_{jk}\frac{1}{s}\left(I_{ij}-\hat{T}_{i}\left(s\right)\delta_{ij}\right)\left(I-\hat{T}\left(s\right)\right)_{jk}^{-1}n_{k}\left(0\right)\,.
	\label{mw_scalar} 
\end{align} 
In terms of vectors and matrices, this is written
\begin{align} 
	\hat{{n}}\left(s\right) 
	& =\frac{1}{s}\left(I-\hat{D}_{T}\left(s\right)\right)\left(I-\hat{T}\left(s\right)\right)^{-1}{n}\left(0\right)\,,
	\label{eq:montroll-weiss} 
\end{align} 
where the diagonal matrix $\hat{D}_{T}$ is defined as 
\begin{equation}
	\left(\hat{D}_{T}\right)_{ij}\left(s\right)=\hat{T}_{i}\left(s\right)\delta_{ij}\,.\label{eq:D_ij(s)}
\end{equation}

Equation (\ref{eq:montroll-weiss}) is a formal solution in Laplace space for the density of a random walk whose dynamics are governed by the WTDs $\psi_{ij}\left(t\right)$. However, taking the inverse Laplace transform to obtain the random-walker density as a function of time does not in general yield closed form solutions.


\subsection{Integro-differential Master Equation}

The generalized master equation (\ref{eq:montroll-weiss}) is an integro-differential equation that describes the evolution of the system in the time domain. In principle, it can be solved numerically to obtain the random-walker density as a function of time. To do this, we need to use the property $\mathcal{L}\left[\frac{d{n}}{dt}\right]=s\hat{{n}}\left(s\right)-{n}\left(0\right)$ of the Laplace transform, which holds provided that the Laplace transform of ${n}\left(t\right)$ and its derivative exist. As we shall show later, the walker density remains properly normalized. 

We define the matrices $A\equiv\left(I-\hat{D}_{T}\left(s\right)\right)$ and $B\equiv\left(I-\hat{T}\left(s\right)\right)$. We then insert $\hat{{n}}\left(s\right)$ from the generalized Montroll-Weiss equation (\ref{eq:montroll-weiss}) to obtain 
\begin{align*}
	\mathcal{L}\left[\frac{d{n}}{dt}\right] 
	&=s\hat{{n}}\left(s\right)-{n}\left(0\right)\\ 
	&=\left(AB^{-1}-I\right)\left(sBA^{-1}\right)\left(\frac{1}{s}AB^{-1}{n}\left(0\right)\right)\\
	&=\left(A-B\right)sA^{-1}\hat{{n}}\left(s\right)\,. 
\end{align*} 
Using the definitions for $A$ and $B$ and simplifying yields 
\begin{align}
	\mathcal{L}\left[\frac{d{n}}{dt}\right] 
	&=\left(\hat{T}\left(s\right)-\hat{D}_{T}\left(s\right)\right)\frac{s}{I-\hat{D}_{T}\left(s\right)}\hat{{n}}\left(s\right)\nonumber \\ 
	& =\left(\hat{T}\left(s\right)\hat{D}_{T}^{-1}\left(s\right)-I\right)\frac{s\hat{D}_{T}\left(s\right)}{I-\hat{D}_{T}\left(s\right)}\hat{{n}}\left(s\right)\nonumber \\ 
	& =\left(\hat{T}\left(s\right)\hat{D}_{T}^{-1}\left(s\right)-I\right)\hat{K}\left(s\right)\hat{{n}}\left(s\right)\,,
	\label{eq:master-laplace} 
\end{align} 
where we have defined the Laplace transform of the \emph{memory kernel} $K$ as \begin{equation} \hat{K}\left(s\right)\equiv\frac{s\hat{D}_{T}\left(s\right)}{I-\hat{D}_{T}\left(s\right)}\,. \label{eq:K(s)} \end{equation} The memory kernel characterizes the amount of memory in the dynamics \cite{balescu}. 
The memory kernel $K(t)$ is usually a function spanning over a non-vanishing time period.  A notable exception is a Poisson process, for which $K(t)=\delta(t)$. The entries of $\hat{D}_{T}\left(s\right)$ are strictly smaller than unity for all finite $s$ because $\hat{T}_j\left(s\right)=\int_0^\infty e^{-st} T_j\left(t\right)dt < \int_0^\infty T_j\left(t\right)dt=1$ (as we shall show in Section \ref{sub:steps-infinite-time}). Hence, the term $I-\hat{D}_{T}\left(s\right)$ is always invertible.


Taking the inverse Laplace transform of (\ref{eq:master-laplace}) leads to the generalized master equation 
\begin{equation}
	\frac{d{n}}{dt}=\left(T\left(t\right)*\mathcal{L}^{-1}\left\{\hat{D}_{T}^{-1}\left(s\right)\right\}-\delta\left(t\right)\right)*K\left(t\right)*{n}\left(t\right)\,, 
	\label{eq:master} 
\end{equation}
where $\mathcal{L}^{-1}$ denotes the inverse Laplace transform and $f*g=\int_{0}^{t}d\tau f\left(t-\tau\right)g\left(\tau\right)$ denotes convolution with respect to time. Unfortunately, mathematical analysis of (\ref{eq:master}) is difficult. However, as we will exploit later, it is often significantly easier to analyze its Laplace-space equivalent (\ref{eq:montroll-weiss}). 

Note that $\hat{D}_{T}\left(s\right)$ can in principle be singular but considering $\hat{D}_{T}^{-1}\left(s\right)\hat{K}\left(s\right)=\frac{s}{I-\hat{D}_{T}\left(s\right)}$ as one term resolves this problem.




\subsection{Conservation of Probability}

The total number of random walkers is a conserved quantity, so $\frac{d}{dt}\sum_{i}n_{i}\left(t\right)=0$. It is not easy to verify this directly in the time domain using (\ref{eq:master}). Thankfully, it is sufficient to show that the Laplace transform of the rate of change of the number of walkers vanishes because $\mathcal{L}^{-1}\left\{ 0\right\} =0$. The inverse Laplace transform is unique except for null functions which are defined by $\int_0^a\mathcal{N}\left( t \right)dt=0$ for all $a>0$. By assuming that the WTDs are piecewise continuous we exclude null functions and the inverse Laplace transform is guaranteed to be unique \cite{dyke1999introduction,spiegel1992laplace}.



Equation (\ref{eq:master-laplace}) gives the rate of change of the total walker density in Laplace space:
\begin{align*} &
\sum_{ij}\left[\left(\hat{T}\left(s\right)\hat{D}_{T}^{-1}\left(s\right)-I\right)\right]_{ij} \left[\hat{K}\left(s\right)\hat{{n}}\left(s\right)\right]_j\\
	&=  \sum_{ikj}\left(\hat{T}_{ik}\left(s\right)\left[\hat{D}_{T}^{-1}\left(s\right)\right]_{kj}-\delta_{ik}\delta_{kj}\right) \left[\hat{K}\left(s\right)\hat{{n}}\left(s\right)\right]_j\\
	&= \sum_{ikj}\left(\hat{T}_{ik}\left(s\right)\frac{1}{\hat{T}_{k}\left(s\right)}-\delta_{ik}\right) \delta_{kj} \left[\hat{K}\left(s\right)\hat{{n}}\left(s\right)\right]_j\\
	&= \sum_{k}\left(\frac{1}{\hat{T}_{k}\left(s\right)}\left[\sum_{i}\hat{T}_{ik}\left(s\right)\right]-1\right) \left[\hat{K}\left(s\right)\hat{{n}}\left(s\right)\right]_k\\
	&= 0 \,, 
\end{align*} 
which confirms that the rate of change of the total walker density vanishes.


\subsection{Making Steps in Infinite Time} \label{sub:steps-infinite-time}





In this paper, we have assumed that the underlying graph $\mathcal{G}$ of potential edges is strongly connected, which implies that a transition from $j$ to some other node is guaranteed to occur eventually if one allows infinite time. We thus expect that $\int_{0}^{\infty}T_{j}\left(t\right)dt=1$, as $T_{j}\left(t\right)$ is the PDF to make any transition from $j$. We use equations (\ref{eq:T_ij(t)-differential-form}) and (\ref{eq:T_i(t)}) to obtain
\begin{align*} 
	T_{j}\left(t\right) =\sum_{i=1}^{N}T_{ij}\left(t\right) 
	& =-\sum_{i=1}^{N}\left(\frac{d\chi_{ij}\left(t\right)}{dt}\times\prod_{k\neq i}\chi_{kj}\left(t\right)\right)\\ 
	& =-\frac{d}{dt}\left(\prod_{i=1}^{N}\chi_{ij}\left(t\right)\right)\,.
\end{align*} 
Integrating over the entire time domain yields 
\begin{align*}
	\int_{0}^{\infty}T_{j}\left(t\right)dt 
	& =-\int_{0}^{\infty}dt\frac{d}{dt}\left(\prod_{i=1}^{N}\chi_{ij}\left(t\right)\right)\\
	& =-\left.\left(\prod_{i=1}^{N}\chi_{ij}\left(t\right)\right)\right|_{t=0}^{\infty} =1\,, 
\end{align*} 
because $\chi_{ij}\left(0\right)=1$ and $\chi_{ij}\left(\infty\right)=0$ when the edge $j\rightarrow i$ exists in the underlying graph, whereas $\chi_{ij}\left(0\right)=1$ and $\chi_{ij}\left(\infty\right)=1$ otherwise.




\subsection{Poisson Limit}

Assume that edge dynamics are Poisson processes and that the WTDs are exponential distributions \cite{stewart2009probability} with a characteristic rate $\lambda_{ij}$ for the transition $j\rightarrow i$: 
\begin{align}
	\psi_{ij}\left(t\right)=\lambda_{ij}e^{-\lambda_{ij}t}\,. 
	\label{lambda}
\end{align} 

Equations (\ref{eq:T_ij(t)}) and (\ref{eq:T_i(t)}) then imply that 
\begin{align}
	\label{ttt} T_{ij}\left(t\right) &=  \lambda_{ij}e^{-\lambda_{ij}t}\prod_{l\neq i}\left(1-\int_{0}^{t}\lambda_{lj}e^{-\lambda_{lj}t'}dt'\right)\\ 
	&=  \lambda_{ij}e^{-\lambda_{ij}t}\prod_{l\neq i}e^{-\lambda_{lj}t} = \lambda_{ij}e^{-\Lambda_{j}t}\,, 
\end{align} 
and 
\[
	T_{j}\left(t\right)= \sum_{i=1}^{N}\lambda_{ij}e^{-\Lambda_{j}t} =\Lambda_{j}e^{-\Lambda_{j}t}\,, 
\]
where the aggregate transition rate from $j$ is defined as $\Lambda_{j}\equiv\sum_{i=1}^{N}\lambda_{ij}$. The Laplace transform of $T_{j}\left(t\right)$ becomes 
\[
\hat{T}_{j}\left(s\right)=\frac{\Lambda_{j}}{\Lambda_{j}+s}\,. 
\]

Equation (\ref{eq:D_ij(s)}) then becomes 
\[
	\left[\hat{D}_{T}\left(s\right)\right]_{ij} =\frac{\Lambda_{j}}{\Lambda_{j}+s}\delta_{ij}\,, 
\]
which in turn yields
\begin{equation} 
	\left[\hat{D}_{T}^{-1}\left(s\right)\right]_{ij}=\left(1+\frac{s}{\Lambda_{j}}\right)\delta_{ij}\,. 
	\label{dd} 
\end{equation} 
Equation (\ref{eq:K(s)}) becomes 
\begin{equation} 
	\label{kk} 
	\hat{K}_{ij}\left(s\right)  
	=\frac{s\frac{\Lambda_{j}}{\Lambda_{j}+s}}{1-\frac{\Lambda_{j}}{\Lambda_{j}+s}}\delta_{ij}
	=\frac{s\Lambda_{j}}{\Lambda_{j}+s-\Lambda_{j}}\delta_{ij}
	=\Lambda_{j}\delta_{ij}\,. 
\end{equation} 
Taking the inverse Laplace transform of (\ref{dd}) yields 
\begin{equation} 
	\label{ddd}
	\mathcal{L}^{-1}\left[\hat{D}_{ij}^{-1}\left(s\right)\right]
	=\left(\delta\left(t\right)+\frac{\delta'\left(t\right)}{\Lambda_{j}}\right)\delta_{ij}\,,
\end{equation} 
and taking the inverse Laplace transform of (\ref{kk}) gives
\begin{equation} 
	\label{kkk} 
	 K_{ij}\left(t\right)=\Lambda_{j}\delta\left(t\right)\delta_{ij}\,. 
\end{equation}

We insert equations (\ref{ttt}), (\ref{ddd}), and (\ref{kkk}) into the generalized master equation (\ref{eq:master}) and use the properties of the Kronecker delta $\delta_{ij}$ and Dirac delta function $\delta(t)$ to obtain 
\begin{equation}
	\frac{dn_{i}}{dt}+\Lambda_{i}n_{i}=\sum_{\mu}\lambda_{i\mu}{e^{-\Lambda_{\mu}t}*\left[\Lambda_{\mu}\delta\left(t\right)+\delta'\left(t\right)\right]}*n_{\mu}\left(t\right)\,.
	\label{eq:almost-rate} 
\end{equation} 
Note that 
\begin{align*} &
	e^{-\Lambda_{\mu}t}*\left[\Lambda_{\mu}\delta\left(t\right)+\delta'\left(t\right)\right]\\
	&= \Lambda_{\mu}\int_{0}^{t}d\tau e^{-\Lambda_{\mu}\left(t-\tau\right)}\delta\left(\tau\right)+\int_{0}^{t}d\tau
e^{-\Lambda_{\mu}\left(t-\tau\right)}\delta'\left(\tau\right)\\ 
	&=\delta\left(t\right)\,, 
\end{align*} 
where the integration of the convolution with $\delta'\left(t\right)$ is performed using integration by parts. Inserting this result into equation (\ref{eq:almost-rate}) yields the rate equation \cite{lambiotte} 
\begin{equation}
	\frac{dn_{i}}{dt}=\sum_{j}\lambda_{ij}n_{j}\left(t\right) - \Lambda_{i}n_{i}\left(t\right)\,, \label{eq:rate} 
\end{equation} 
driven by the combinatorial Laplacian $L_{ij}=\lambda_{ij} - \Lambda_{i} \delta_{ij}$ of a weighted network defined by the adjacency matrix $\lambda_{ij}$. This expression thus shows that a random walk on the dynamical graph driven by (\ref{lambda}) is equivalent to a Poisson continuous-time random walk on a static network, which is constructed in the usual manner by counting the number of times edges appear between each pair of nodes.  In other words, the rate is equal to the number of observed appearances of an edge divided by the duration of the observation. 


\section{Asymptotic Behavior}


\subsection{Steady-State Solutions} \label{sec:sss}



We expect the walker distribution to settle into a unique steady-state solution as $t \rightarrow \infty$ if any node can be reached from any other node. By the so-called final value theorem \cite{jaeger1969introduction}, the steady-state walker density ${p}$ is 
\[
	{p}=\lim_{t\rightarrow\infty}{n}\left(t\right)=\lim_{s\rightarrow0}s\hat{{n}}\left(s\right)\,.
\] 
The final value theorem requires that $\hat{{n}}\left(s\right)$ does not have poles in the right half of the complex $s$-plane or on the imginary axis (except possibly at the origin).
As we shall show later, $s\hat{{n}}\left(s\right)$ has a pole at $s=0$, and only this pole contributes to the steady-state solution \cite{jaeger1969introduction}. 



From equation (\ref{eq:montroll-weiss}), one obtains 
\begin{align*} 
	p & =\lim_{s\rightarrow0}\left(I-\hat{D}_{T}\left(s\right)\right)\left(I-\hat{T}\left(s\right)\right)^{-1}{n}\left(0\right)\\
	& =\mathbb{M}{n}\left(0\right)\,, 
\end{align*} 
where the matrix
\[
	\mathbb{M}\equiv\lim_{s\rightarrow0}\left(I-\hat{D}\left(s\right)\right)\left(I-\hat{T}\left(s\right)\right)^{-1}
\]
maps the initial state to the final state. In the limit $s\rightarrow0$, one can expand the exponential in the definition of the Laplace transform to first order: 
\begin{align*}
	\left[\hat{D}_{T}\left(s\right)\right]_{ij}
	&=\int_{0}^{\infty}\left[1-st+\mathcal{O}\left(s^{2}\right)\right]T_{j}\left(t\right)\delta_{ij}dt\\
	& =\left[1-s\langle t_{j}\rangle+\mathcal{O}\left(s^{2}\right)\right]\delta_{ij}\\ & =I-sD_{\langle t\rangle}+\mathcal{O}\left(s^{2}\right)\,, 
\end{align*} 
where $\langle t_{j}\rangle=\int_{0}^{\infty}tT_{j}\left(t\right)dt$ is the mean time spent on node $j$ and we have defined the diagonal matrix $D_{\langle t\rangle}=\langle t_{j}\rangle\delta_{ij}$. One can also use the approximation 
\begin{align}
	\label{coot} 
	\hat{T}_{ij} 
	& =\int_{0}^{\infty}\left[1-st+\mathcal{O}\left(s^{2}\right)\right]T_{ij}\left(t\right)dt \notag \\ 
	& =\mathbb{T}_{ij}\left(1-s\int_{0}^{\infty}t\frac{T_{ij}\left(t\right)}{\mathbb{T}_{ij}}dt+\mathcal{O}\left(s^{2}\right)\right) \notag \\ 
	& =\mathbb{T}_{ij}\left[1-s\langle t_{ij}\rangle+\mathcal{O}\left(s^{2}\right)\right]\,, 
\end{align} 
where $\langle t_{ij}\rangle$ is the mean time before making a step $j\rightarrow i$.
\[
	\mathbb{T}_{ij} \equiv \int_0^\infty T_{ij}(t) dt 
\]
is the probability of making a step $j\rightarrow i$ and is a measure of the relative importance of the edge. $\mathbb{T}$ is called the \emph{effective transition matrix}. One can write (\ref{coot}) in matrix form as 
\[
	\hat{T} = \mathbb{T}-s\mathbb{T}\circ\langle t\rangle+\mathcal{O}\left(s^{2}\right)\,, 
\]
where $\circ$ denotes the Hadamard component-wise product.





The operator $\mathbb{M}$ becomes 
\[
	\mathbb{M}=\lim_{s\rightarrow0}\left[sD_{\langle t\rangle}\left(I-\mathbb{T}-s\mathbb{T}\circ\langle t\rangle\right)^{-1}\right]+\mathcal{O}\left(s^{2}\right)\,, 
\] 
and it is expected to project any initial condition onto a unique one-dimensional vector if the underlying graph $\mathcal{G}$ consists of one strongly connected component. The steady-state solution is thus given by the leading eigenvector (i.e., the eigenvector corresponding to the maximum eigenvalue) of the projection operator $\mathbb{M}$. In practice, it is easier to obtain the least dominant eigenvector (i.e., the eigenvector whose corresponding eigenvalue is the smallest in absolute value) of the inverse of the projection operator 
\begin{align*}
	\mathbb{M}^{-1} 
	& =\lim_{s\rightarrow0}\frac{1}{s}\left[I-\mathbb{T}-s\mathbb{T}\circ\langle t\rangle\right] D_{\langle t\rangle}^{-1}\\ 
	&=\lim_{s\rightarrow0}\left[\frac{1}{s}\left(I-\mathbb{T}\right)D_{\langle t\rangle}^{-1}-\left(\mathbb{T}\circ\langle t\rangle\right)D_{\langle t\rangle}^{-1}\right]\,. 
\end{align*}
In the limit $s\rightarrow 0$, the eigenvectors of $\mathbb{M}^{-1}$ tend to the eigenvectors of the matrix $C\equiv\left(I-\mathbb{T}\right)D_{\langle t\rangle}^{-1}$ because the second term in the above equation becomes negligible in comparison to the first. Thus, finding the least dominant eigenvector of $C$ is equivalent to finding the dominant eigenvector of $\mathbb{M}$ in the limit $s\rightarrow 0$. Note that $D_{\langle t\rangle}$ is invertible unless the mean time spent on any node $\langle t_j\rangle$ is zero. This would imply that any walker that arrives at $j$ immediately makes a step to another node $i$ such that the labels $i$ and $j$ actually refer to the same node. $D_{\langle t\rangle}$ is thus always invertible in practice.



The matrix  $\mathbb{T}$ is a stochastic matrix because its columns are normalized (as we have shown in Section \ref{sub:steps-infinite-time}). In particular, $\mathbb{T}$ has a single eigenvector, which we denote by $x$, whose eigenvalue is equal to unity if $\mathcal{G}$ is strongly connected \cite{stewart2009probability}--- i.e., $\mathbb{T} x=x$.  This implies that the steady-state walker density $p$ is given by
\begin{equation}
	\label{soloo}
	{p}=D_{\langle t\rangle}{x} 
\end{equation} 
as it is the least dominant eigenvector of $C$:
\begin{align*} 
	C{p} &=\left(I-\mathbb{T}\right)D_{\langle t\rangle}^{-1}D_{\langle t\rangle}{x}\\ 
& =I{x}-\mathbb{T}{x} =0\,. 
\end{align*} 

Given the complexity of the stochastic process and the integro-differential nature of its associated master equation, it is remarkable that one can write down such an exact analytical expression. This equilibrium solution takes a particularly simple form, as it is calculated from the dominant eigenvector of the combinatorial Laplacian associated with the effective transition matrix $\mathbb{T}$. The time spent on node $i$ is thus given by the frequency to arrive on it (which is obtained by the Markov chain associated to $\mathbb{T}$) multiplied by the waiting time $\langle t_{i}\rangle$ spent on it. This solution can be computed easily even for very large graphs, as deriving $\mathbb{T}$ from $\psi$ is straightforward and the leading eigenvector of a large matrix can be obtained through standard, efficient techniques (e.g., the power method) \footnote{Note that if $\mathcal{G}$ is made of $k$ strongly connected components, then there are $k$ different eigenvectors with unit eigenvalues.  This would then lead to $k$ distinct steady-state solutions rather than the unique one in the strongly connected case.}. 


It remains to be shown that $s\hat{{n}}\left(s\right)$ has a pole at $s=0$, which would guarantee that the requirements of the final value theorem are satisfied. From equation (\ref{eq:montroll-weiss}), it follows that $\hat{{n}}\left(s\right)$ has poles whenever  $I-\hat{T}(s)$ is singular. Using (\ref{coot}), we find  in the limit $s\rightarrow 0$ that
\begin{align*}
        \left(I-\hat{T}(s)\right)x&=\left(I-\mathbb{T}-s\mathbb{T}\circ\langle
        t\rangle\right) x+\mathcal{O}\left(s^{2}\right)\\
        &=Ix-\mathbb{T}x-s\mathbb{T}\circ\langle t\rangle x+\mathcal{O}\left(s^{2}\right)\\
        &=-s\mathbb{T}\circ\langle t\rangle x+\mathcal{O}\left(s^{2}\right).
\end{align*}
Hence, $x$ is an eigenvector of $I-\hat{T}(s)$ with eigenvalue $0$. Because $\det\left(I-\hat{T}(s)\right)$ is equal to the product of eigenvalues, $I-\hat{T}(s)$ becomes singular as $s\rightarrow 0$.



\subsection{Poisson Versus Non-Poisson Processes}

We now discuss some of the differences arising from the non-Poisson nature of the process versus the Poisson situation that is usually considered. For simplicity, we illustrate these differences using an undirected network, for which $\psi_{ij}=\psi_{ji}$.

For the Poisson process (\ref{lambda}), it is straightforward to show from equation (\ref{ttt}) that the effective transition matrix $\mathbb{T}$ satisfies 
\begin{equation}
\label{popopo}
	\mathbb{T}_{ij}= \frac{\lambda_{ij}}{\sum_{i=1}^{N}\lambda_{ij}}\,,
\end{equation}
where $\lambda_{ij}$ is the adjacency matrix of an undirected and static weighted network, where the weights are the rates at which edges appear in the dynamical network. The stationary solution is a uniform vector, which is expected, as it is the dominant eigenvector of the Laplacian matrix in (\ref{eq:rate}).



The case of a non-Poisson process differs significantly from this idealized scenario in at least three ways. First, $\mathbb{T}$ is no longer in general the transition matrix of an undirected network, even in the symmetric case $\psi_{ij}=\psi_{ji}$ that we consider.  Evaluating the stationary solution thus requires one to calculate the dominant eigenvector ${x}$ of $\mathbb{T}$. Second, there is no reason to expect that the stationary solution ${p}$ is uniform; indeed, it is not uniform in general. Finally, ${p}$ is a stationary solution only in the limit $t \rightarrow \infty$ and not also for intermediate times, as we used an expansion for small $s$ throughout section \ref{sec:sss} to derive its expression. This behavior originates from the time-dependent nature of the stochastic process, as one can see because of the integral over time in the generalized master equation (\ref{eq:master}). This has interesting implications. For example, a process starting with the stationary solution ${p}$ as an initial condition exhibits transient deviations to stationarity before returning to its initial condition as $t \rightarrow \infty$. In contrast, if the dynamics on the graph are Poisson, then the steady-state solution is time-independent. In other words, once the system has reached its steady-state solution, it will stay there.

\begin{figure} 
\begin{centering}
\includegraphics[width=0.35\textwidth]{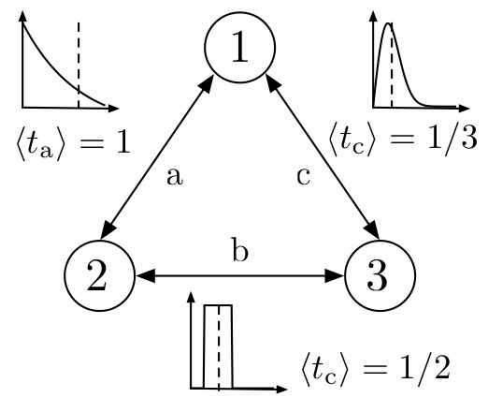} \par\end{centering}
\caption{\label{fig:ss-example} An undirected network with $N=3$ nodes and no self-loops. The waiting time distributions used for the edges $a$, $b$, and $c$ are exponential, $\chi^{2}$, and Rayleigh, with means $\langle t_a\rangle$, $\langle t_b\rangle$ and $\langle t_c\rangle$ respectively.} 
\end{figure}

\begin{figure} \begin{centering}
\includegraphics[width=0.4\textwidth]{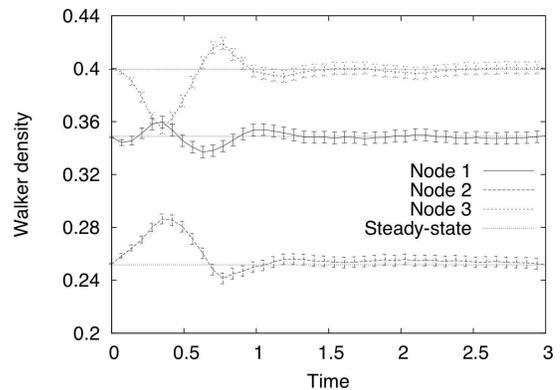}
\par\end{centering} \caption{\label{numericalstability} Random-walker density on each node of the graph illustrated in Fig.~\ref{fig:ss-example} as a function of time when the initial condition is the steady-state solution. Due to the time-dependent nature of the dynamics, the system exhibits transient dynamics before returning to its steady-state solution. The error bars again give a confidence interval of 5 standard deviations.
} 
\end{figure}

To illustrate how the nature of the WTDs affects dynamics, consider the toy example of a fully-connected graph with three nodes. Suppose that the WTDs for the processes occurring on the nodes have different functional types and different characteristic times (see Fig.~\ref{fig:ss-example}). We compare this non-Poisson case to a Poisson process with the same average rates $\lambda_{ij}$. The associated matrix, which corresponds to the adjacency matrix of (\ref{popopo}), is 
\[ 
	\lambda=\left(\begin{array}{ccc} 
	0 & 1 & 3\\ 
	1 & 0 & 2\\ 
	3 & 2 & 0 \end{array}\right)\,. 
\] 
We illustrate the differences between the two processes in Fig.~\ref{fig:analytic-numerical}, where we perform numerical simulations of the random walk (see the Appendix for computational details) with all walkers located at node $1$ initially, and we plot the temporal evolution of the walker density. The system relaxes towards a stationary solution in both cases, but this stationary solution clearly depends on the nature of the WTDs, as walkers tend to be overrepresented on node $1$ for the non-Poisson dynamics. Figure~\ref{numericalstability} clearly conveys the importance of the time-dependence of the process and the fact that its steady-state solution is truly steady only at long times.






\begin{figure} \begin{centering}
\includegraphics[width=0.4\textwidth]{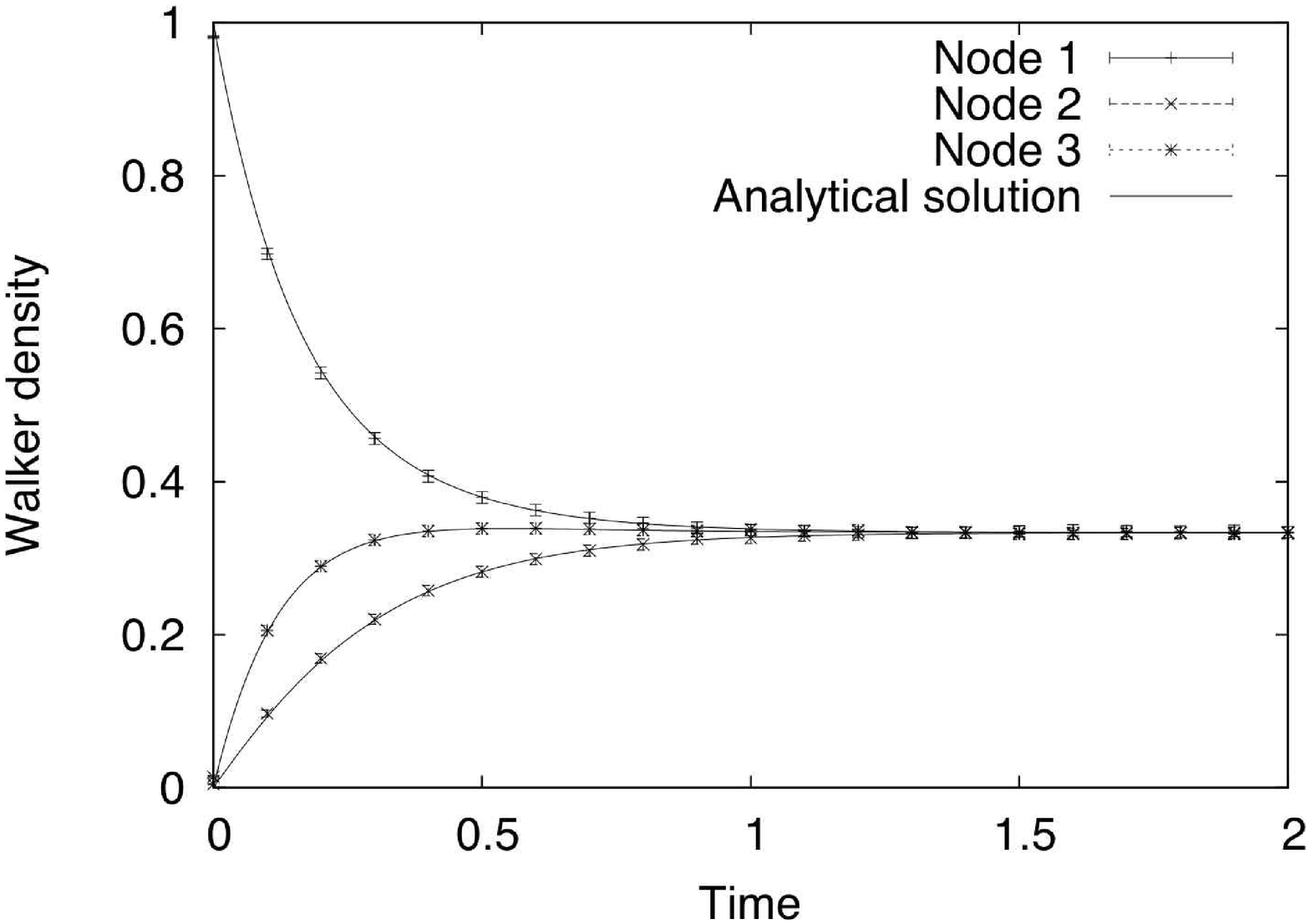}
\includegraphics[width=0.4\textwidth]{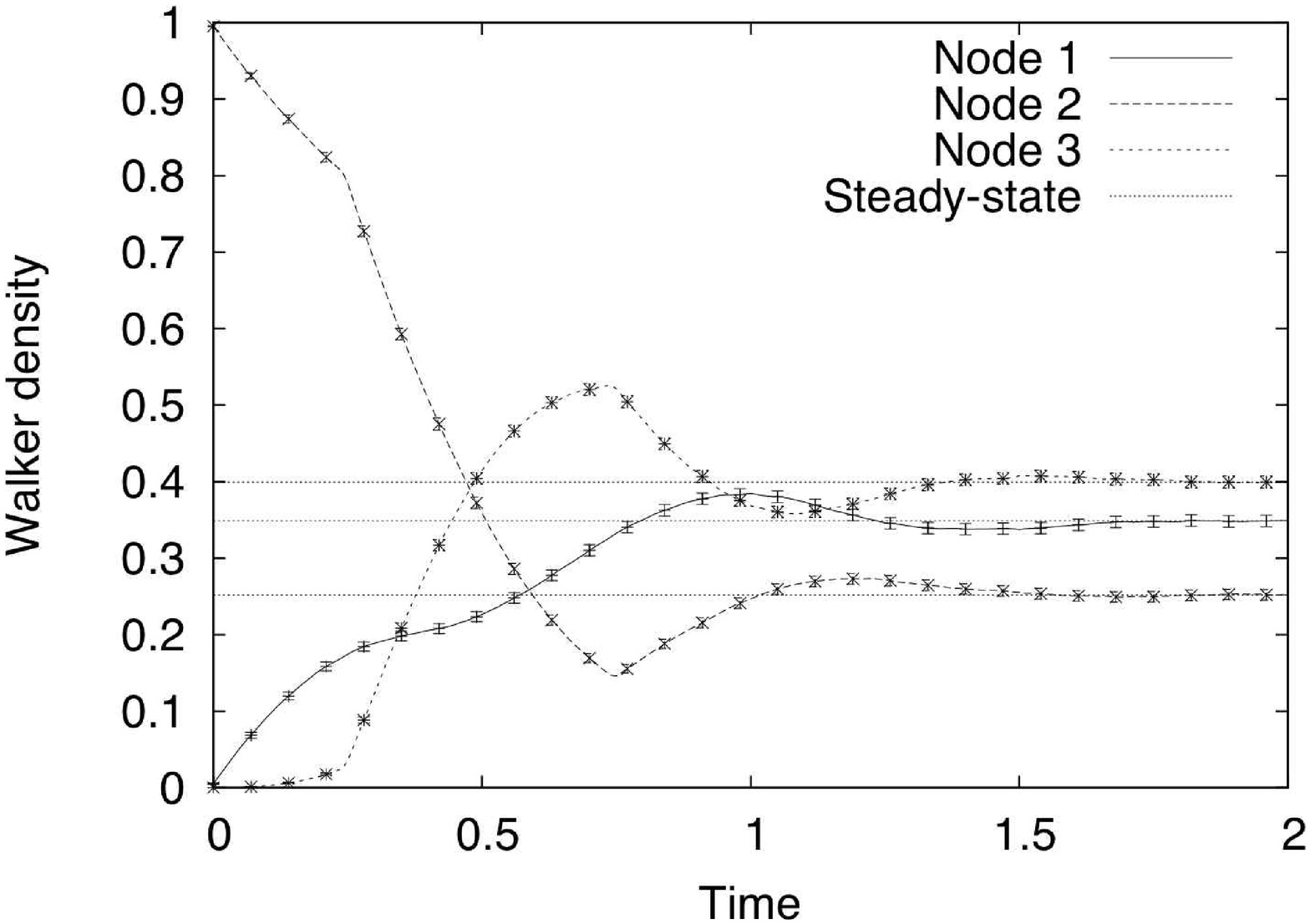} \par\end{centering}
\caption{\label{fig:analytic-numerical} Random-walker density on each node as a function of time obtained from numerical simulations for the Poisson (upper panel) and non-Poisson (lower panel) processes. In the former case, we also plot the analytical solution of the rate equation. The error bars give a confidence interval of 5 standard deviations. 
For the non-Poisson example, we obtained the steady-state walker density from equation (\ref{soloo}).The relaxation towards stationarity exhibits kinks in the dynamics that originate from the non-continuity of the waiting-time distributions. 
} 
\end{figure}

\section{Discussion}

The overwhelming majority of research that considers dynamics on networks implicitly includes a Poisson assumption. For instance, this is the case when one builds an aggregated network from empirical data and assigns only a single number (i.e., a weight) to edges to summarize their dynamics. In that framework, edge weight is understood as a rate of interaction between nodes, so such a choice would imply that events occur in an uncorrelated fashion. The main purpose of the present work is to consider a mathematical framework that goes beyond this oversimplification and thereby propose a compromise between abstract but unrealistic models and data-driven but non-mathematical approaches for studying temporal networks. In the proposed framework, temporal networks are viewed as sequences of realizations of random networks in which present events depend on past ones. The temporal and structural organization of a system is summarized by a matrix of waiting time distributions (WTDs) that characterize the pattern of activation of edges between nodes. In this paper, we have derived a generalized master equation for a random walk associated with this process, obtained an expression for its stationary solution, and used a toy network to highlight the importance of the shape of WTDs.

The present paper is a step towards a mathematical description of temporal networks, and it suggests several research directions. An important step will be thorough consideration of empirical data, such as human communication patterns, to help design appropriate WTDs and to study how their shape affects equilibrium solutions.  This step is related to the development of network diagnostics that properly take into account the temporal dynamics of complex systems \cite{temporalmetrics,grindrod,pathlengthtemporal,Mucha}, whose behavior arises not only from their structural patterns but also from their dynamics. The investigation of complex systems thus requires the development of diagnostics that consider both structure and dynamics \cite{lambiotte}. The fact that most existing network diagnostics account only for the presence, weight, and direction of edges and assume that they are present for all time presupposes an underlying Poisson process, which can yield a fundamentally incorrect representation of a system.  Indeed, assuming a Poisson waiting time can be as unrealistic as assuming that the topology of a network is organized like an Erd\"os-R\'enyi random graph. It is thus crucial to develop network concepts that account for nontrivial temporal dynamics. This includes centrality measures (like Pagerank \cite{pagerank}), which measure the importance of network components in various ways, and the components of the steady-state solution vector derived in this paper provide an example of one such diagnostic. Another example is community-detection methods based on the idea that non-Poisson random walkers are trapped for long times in good communities \cite{rosvall,delvenne}, though of course there is much more that one can do. Other interesting research directions include the development of a sound mathematical framework for various dynamical processes, such as synchronization or epidemic spreading \cite{new2} and the study of random walks with alternative ways to build the matrix of waiting times. For example,  the dynamics could be driven by the normalized Laplacian in the Poisson limit.


\begin{acknowledgments} 

We would like to thank T. Carletti, J.-C. Delvenne, P. Mucha, M. Rosvall, and J. Saram\"aki for fruitful discussions. 

\end{acknowledgments}


\newpage

\appendix \section{Numerical Simulation}

We implemented the numerical simulation in C\# using the {\tt Math.NET} library to simplify the generation of random numbers from various distributions. 


The approach in this paper requires computing the probability of finding a walker on a certain node. To approximate this numerically, we simulate $10^5$ random walks for each scenario and average over the realizations of the walk in order to estimate the walker density numerically.




\subsection{Generating Trajectories}

The trajectory of a walker is generated by making consecutive steps until the maximum simulation time $t_{\mbox{max}}$ is reached. See Algorithm \ref{alg:trajectory}.

\begin{algorithm} 
\begin{algorithmic}[1] 
\STATE history := new History() \{Create a store to keep track of steps.\} 
\STATE history.CurrentNode := \textbf{input}(startNode) \{Get the initial node.\} 
\WHILE{(history.Time$\leq t_{\mbox{max}}$)} 
	\STATE currentEdges := \textbf{from} edge \textbf{in} $\mathcal{G}$.Edges \textbf{where} edge.SourceIndex = history.CurrentNode.Index) \textbf{select} edge \{Select a list of edges over which a walker can leave the current node.\} 
	\STATE step := \textbf{new} Step() \{Create a new record.\}
	\STATE step.Delta := $\infty$ \{Declare a minimum wait.\} 
	\FOR{\textbf{each} edge \textbf{in} currentEdges} 
		\STATE sample := edge.Distribution.GetSample() \{Obtain a sample from the WTD of the edge.\} 
		\IF {(sample $<$ step.Delta) \{Does the step along this edge occur before all previous edges?\}} 
			\STATE step.Delta := sample \{Assign the new minimum wait.\} 
			\STATE step.Edge := edge \{Store the associated edge.\} 
		\ENDIF 
	\ENDFOR 
	\STATE history.Add(step) \{Add the step to the history.\} 
\ENDWHILE \end{algorithmic}
\caption{\label{alg:trajectory} Generation of the trajectory of a random walker.} 
\end{algorithm}


\subsection{Averaging Over Different Realizations}

\begin{figure} \begin{centering}
\includegraphics[width=0.4\textwidth]{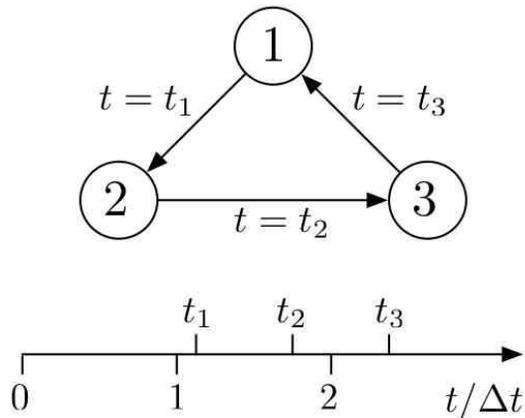} \par\end{centering}
\caption{\label{fig:walk}A random walker makes three steps on a graph of three
nodes at irregular intervals.} \end{figure}

Steps in each of the $10^5$ different realizations are made at irregular intervals. We consider a set of $k$ finite intervals of width $\Delta t = 0.01$ such that $k\Delta t=t_{\mbox{max}}$. The probability to find a random walker on node $j$ during a certain interval is then given by the mean over all realizations of the fraction of the time spent on $j$ during this interval. Consider a single realization of a random walk on a graph of three nodes (see Fig.~\ref{fig:walk}). In Table \ref{tab:frac-times}, we list the fractions of time spent on each of the nodes in the contribution to the average.  


\begin{table} \begin{centering} \begin{tabular}{|c|c|c|c|} 
\hline Interval & Node 1 & Node 2 & Node 3\tabularnewline 
\hline \hline 1 & 1 & 0 & 0\tabularnewline 
\hline 2 & 0.125 & 0.625 & 0.25\tabularnewline 
\hline 3 & 0.625 & 0 & 0.375\tabularnewline 
\hline \end{tabular} \par\end{centering}
\caption{\label{tab:frac-times}Fractions of time spent on each node during the first three intervals.} \end{table}


\end{document}